# Moderate Degree of Input Negative Entropy Flow and Decrease of Entropy in Astrophysics, Biology, Psychology and Social Systems

Yi-Fang Chang
Department of Physics, Yunnan University, Kunming, 650091, China
(e-mail: yifangchang1030@hotmail.com)

**Abstract**: Thermodynamics have been applied to astronomy, biology, psychology, some social systems and so on. But, various evolutions from astronomy to biology and social systems cannot be only increase of entropy. When fluctuations are magnified due to internal interactions, the statistical independence and the second law of the thermodynamics are not hold. The existence of internal interactions is necessary condition of decrease of entropy in isolated system. We calculate quantitatively the entropy of plasma. Then we discuss the thermodynamics of biology, and obtain a mathematical expression on moderate degree of input negative entropy flow, which is a universal scientific law. Further, the thermodynamics of physiology and psychology, and the thought field are introduced. Qigong and various religious practices are related to these states of order, in which decrease of entropy is shown due to internal interactions of the isolated systems. Finally we discuss possible decrease of entropy in some social systems.

## 1.Possible decrease of entropy in astrophysics

Thermodynamics in astrophysics includes one of black hole [1,2]. Kaburaki, et al., discussed thermodynamic stability of isolated Kerr black holes with respect to axisymmetric perturbations [3]. Hochberg, et al., calculated that the thermodynamical entropy $\Delta S$ is positive in the semiclassical theory of black holes and radiation, in which w=2M/r, $1 \leq w \leq w_0 = 2M/r_0$, here M is the mass of the black hole, $\partial(\Delta S)/\partial w$ vanishes at the horizon w=1 [4]. Carter and Neupane studied the thermodynamic and gravitational stability of Kerr anti-de Sitter black holes in five and higher dimensions [5].

Saslaw and Hamilton discussed thermodynamics and galaxy clustering in nonlinear theory of high order correlations [6]. Itoh, et al., compared between thermodynamic theory and n-body for gravitational clustering of galaxies [7]. Mihalas, et al., studied the equation of state for stellar envelopes for thermodynamic quantities [8]. Muller, et al., obtained the entanglement entropy $S_{ent} = 0.30 r_H^2 / a^2$ in curved spacetimes with event horizons [9]. Saslaw and Fang investigated three fundamental aspects of the thermodynamic description of the cosmological many-body problem [10]. Boss searched systematically evolution of the solar nebula [11], in particular, the gas giant protoplanet formation in disk instability with varied thermodynamics.

If the total Universe tends to maximum entropy, it will be the heat death, which is the most controversial question for thermodynamics. But, we cannot assume that the total Universe is open. In fact the basis of thermodynamics is the statistics, in which a basic principle is statistical independence [12]. It shows that various interactions among these subsystems should not be considered. When various internal interactions cannot be neglected, a state with smaller entropy can appear, for example, self-organized structure, etc. So decrease of entropy in an isolated system is possible [13-15], in particular, for attractive process, system entropy, nonlinear interactions, chemical reactions, quantum systems, microstructure and so on [16-18].

For any isolated system we proposed a generalized formula [15]:

$$dS = dS^a + dS^i, \qquad (1)$$

where $dS^a$ is an additive part of entropy and is always positive, and $dS^i$ is an interacting part of entropy and can be positive or negative. From this we obtained the sufficient-necessary condition of decrease of entropy in isolated system [18].

In some cases, internal interactions are very important. For example, many protons mix with electrons to form hydrogen atoms, a pair of positive and negative ions forms an atom, and various neutralization reactions between acids and alkalis form different salts. These far-equilibrium nonlinear processes form some new self-organized structures due to electromagnetic interactions.

Any stable objects and their formations from particles to stars are accompanied with internal interactions inside these objects, which imply a possibility of decrease of entropy [19]. Various evolutionary processes from astronomy to biology and social systems cannot be only increase of entropy.

Almost all matter in interstellar space can be considered to be plasma, an ionized gas consisting of electrons, ions, and neutral atoms or molecules [20]. The pressure P of a nondegenerate gas is

$$P = N_{tot}\frac{kT}{V} = \rho\frac{kT}{\mu}. \qquad (2)$$

For the gas the entropy is

$$S = kN_{tot}\ln\left[\frac{C(kT)^4}{P}\right] = kN_{tot}\ln\left[CP^3\left(\frac{\mu}{\rho}\right)^4\right]. \qquad (3)$$

In an isolated system, $N_{tot}$ is invariant,

$$S - S_0 = kN_{tot}\ln\left[\left(\frac{P}{P_0}\right)^3\left(\frac{\rho_0}{\rho}\right)^4\right]. \qquad (4)$$

For a plasma,

$$P = N_{tot}(1+\alpha)\frac{kT}{V} = (1+\alpha)\frac{kT\rho}{\mu}, \qquad (5)$$

where $\alpha = N_e/N_{tot}$ is the ionicity. Such the entropy of plasma is

$$S - S_0 = 3kN_{tot} \ln\left[\left(\frac{T}{T_0}\right)\left(\frac{\rho_0}{\rho}\right)^{\gamma-1} \frac{1+\alpha}{1+\alpha_0}\right]. \tag{6}$$

where $\gamma = c_P/c_V$, $\gamma \approx 4/3$ for the polyatomic gases. In an isolated system of plasma, the ions will attract electrons to neutral atoms. Since the plasma is often taken to be in local thermodynamic equilibrium [20], and the total mass of an isolated system is constant, temperature T and average density $\rho$ are invariant. Therefore, the final ionicity changes smaller, so that the final entropy decreases [14].

For a non-ideal gas, the thermodynamic potential [21] is

$$\Omega = -PV = -<N>kT\sum_{l=1}^{\infty} B_l(T)\left(\frac{<N>}{V}\right)^{l-1}, \tag{7}$$

where $B_l$ is the so called the virial coefficient. Assume $<N>/V=C$(constant), the entropy is

$$S = -\frac{\partial \Omega}{\partial T} = \frac{\partial}{\partial T}\left\{\frac{kT}{V}\sum_{l=1}^{\infty} C^l B_l(T)\right\} = \frac{k}{V}\sum_{l=1}^{\infty} C^l (B_l + T\frac{\partial B_l}{\partial T}). \tag{8}$$

Let

$$v(q) = 4\varepsilon\left[(\frac{\sigma}{q})^{12} - (\frac{\sigma}{q})^{6}\right], \tag{9}$$

be the Lennard-Jones 6-12 potential [21], then

$$B_2 = b_0 \sum_{n=0}^{\infty} \alpha_n (\frac{1}{T^*})^{(2n+1)/4}, B_3 = b_0^2 \sum_{n=0}^{\infty} \beta_n (\frac{1}{T^*})^{-(n+1)/2}.(T^* = \frac{kT}{\varepsilon}), \tag{10}$$

$$S = \frac{kC}{V}\left\{1 + Cb_0 \sum_{n=0}^{\infty} \alpha_n \frac{3-2n}{4}(T^*)^{-(2n+1)/4} + (Cb_0)^2 \sum_{n=0}^{\infty} \beta_n \frac{n+3}{2}(T^*)^{(n+1)/2}\right\}. \tag{11}$$

If $B_l$ for $l \geq 3$ are neglected,

$$S = \frac{kC}{V}\left\{1 + Cb_0\left[\alpha_0 \frac{3}{4}(T^*)^{-1/4} + \frac{1}{4}\alpha_1(T^*)^{-3/4} - \frac{1}{4}\alpha_2(T^*)^{-5/4} - \frac{3}{4}\alpha_3(T^*)^{-7/4} - ...\right]\right\}. \tag{12}$$

Since $\alpha_0 > 0$ for n=0, and $\alpha_n < 0$ for n>0, and $b_0 = 2\pi\sigma^3/3 > 0$, all terms of the entropy (12) are positive except the $\alpha_1$ term. Therefore, the entropy is positive, and decreases as the temperature rises and $T^{-\mu}$ decreases. This is consistent with the formula dS=dU/T.

The galaxy may be regarded as an isolated system [22]. Based on the virial theorem, the total energy of the system is finite. For an inverse square law force, as in gravitation or electrostatics, the average potential energy is V=-2K [22]. When the stars are formed from nebula, and the galaxies are formed a medium in the Universe, an assembly of molecules is considered, their average kinetic energy K per mole is

$$K = (3/2)(c_P - c_V)T = (3/2)(\gamma - 1)c_V T. \tag{13}$$

$$V = -3(\gamma - 1)U, \tag{14}$$

where $U = c_V T$ is the internal energy. The total energy per mole is

$$E = U + V = -(3\gamma - 4)U = \frac{3\gamma - 4}{3\gamma - 3}V. \tag{15}$$

For $\gamma > 4/3$, E is always negative, and the system is bound. If the system contracts and the potential energy changes by $\Delta V$, then

$$\Delta E = \frac{3\gamma - 4}{3\gamma - 3}\Delta V = -(3\gamma - 4)\Delta U. \tag{16}$$

Hence, the internal energy increases by

$$\Delta U = -\frac{1}{3\gamma - 3}\Delta V, \tag{17}$$

due to a rise in temperature. As the protostar contracts to form a star, it therefore becomes hotter and hotter. We may describe the formation of stars through the compression of cool gas clouds, since, as it collapses, the protostellar cloud becomes hotter and hotter [22]. Thus, the disorder protostellar cloud forms regular stars through self-interactions in the evolutionary process. For the origin of stars, Zhang Bang-gu calculated quantitatively the entropy of contracted gas group, and results show decrease of entropy [23].

The potential (9) agrees well with a very strong repulsive hard-core and a short-range attractive force [21], which just corresponds to the microscopic strong interaction.

Moreover, in astronomy Hawking black hole evaporation theory derives partly reversibility of arrow of time for black hole [24]. Hawking discussed that the thermodynamic arrow would reverse during a contracting phase of the Universe or inside black holes [25]. The white hole [26], which was first suggested in 1964 by I.Novikov and M.Hjellming as a pure mathematical creature [27], is a time-reversed black hole. We proposed that the grand unified theory in particle physics may apply to the supermassive stars, whose energy scale is large enough as to take an infinitely collapsing process, and derived a new model after nucleon-decays, in which a supermassive star will convert nearly all its mass into energy, and become a lepton star that is possibly substable or unstable. According to the model the ultrahigh energy cosmic rays and these puzzles (including quasars and gamma-ray bursts, etc.) in astrophysics at high energy may be explained. The model possesses some properties of the white holes, in which baryon-number is nonconservation. Perhaps, it is a true white hole [28,29].

**2. Moderate degree on input negative entropy flow and possible decrease of entropy in biology**

In 1945, Erwin Schrodinger published a farsighted book about the concept of life [30]. The book was devoted to the relationship between physics and biology. He discussed three problems of fundamental importance for biophysics: the thermodynamic bases of life and the molecular basis of life, while emphasizing those normal biological processes are consistent with the laws of physics. Schrodinger pointed out: An organism feeds with negative entropy. His arguments demonstrate that life possesses lower entropy. Edsall and Gutfreund discussed generally biothermodynamics [31]. W.Ebeling (1985) researched thermodynamics of self-organization and evolution. Kennedy researched biothermodynamics for sustainability of action in ecosystems [32].

Further, biology requires the nonlinear thermodynamics. Ilya Prigogine, et al., developed nonlinear thermodynamics and a dissipative structure theory [33,34]. Schaleger and Long discussed entropies of activation and mechanism of reactions in solution, in which the less negative entropy of activation observed for the hydrolysis of the formate ester was thought to be due to the free rotation of the alkyl group being restricted in going from the trigonal reactant to the tetrahedral transition state [35]. Page and Williams discussed entropy of activation, in which smaller entropy changes from the restriction of internal degrees of freedom occur in cyclization reactions [36]. The electrostriction effect is well exemplified by the dissociation of neutral species such as carboxylic acids where the ionization constant is associated with some 84J/(Kmol) decrease in entropy. Hydroxide ion attack on esters is associated with a large decrease in entropy (Table 5.3 in [36]).

In the dissipative structure theory, the total change of entropy for an open nonequilibrium system is

$$dS = dS_i + dS_e, \qquad (18)$$

where $dS_i \geq 0$ is the entropy production inside the system, and $dS_e$ is the entropy flow, which may be positive or negative, such that $dS_e = dS_e^+ - dS_e^-$. The total entropy can decrease when input entropy flow is negative.

The total entropy is then given as

$$S = S_0 + dS = S_0 + dS_i + dS_e^+ - dS_e^- > 0, \qquad (19)$$

which and the entropy production are always positive. Therefore, the maximum entropy is

$$S_{max} = S_0 + dS_i + dS_e^+ \geq S_0 \geq dS_e^- > 0. \qquad (20)$$

The maximum entropy defines a quantitative range of moderate degree on input negative entropy flow for any open system [37]. Its absolute value is always greater than zero, but the total entropy can never become negative.

In the general goal for input negative entropy flow is: (a) An existing order structure is kept, such that negative entropy flow equal entropy production, and the total entropy is invariance so that $S = S_0$ =constant. And (b) it allows for internal entropy fluctuations, which imply the construction of a new order structure. In the second case, it is common for $dS_e^+ = 0$ and $dS_e^- > dS_i$, so that the total entropy decrease and dS<0.

Under the condition defined in equation (20), an input value of negative entropy flow can be neither excessively large nor small. Excessively small values prohibit the existence of a dissipative structure and do not achieve the threshold value for transformation to a new order structure. Conversely, if the excessively large values are beyond the sustained power of the system itself or the particular circumstances governing the system, it will break various stabilities. Therefore, the moderate degree on input negative entropy flow includes a control of open degree in system and a selection of input time. The input negative entropy flow is determined by the internal conditions of system and is restricted by the external circumstances.

In either case, all living systems are very complex. Their entropies can be neither overly large nor small and the input negative entropy must have a period. For any open system, a rational combination between the input period and the input amount of negative entropy flow is guaranteed for either a stable structure or for the continual transformation to newer ordered structures. These conditions are suitable for any living system. The moderate degree of input negative entropy flow is a universal scientific law. It is suitable for various natural and social systems, and human.

Eigen proposed the hypercyle theory, which discussed self-organization of matter and the evolution of biological macromolecules [38]. It is in order to model prebiotic evolution governed by the Darwinian principles of competition between species and mutations, and leads to a new level of evolution. Here cooperative behaviors are reflected by intrinsically nonlinear reaction mechanisms. The hypercycle theory may find important applications in fields other than biomolecules.

The auto-control mechanism in an isolated system may produce a degree of order. If it does not need the input energy, at least in a given time interval, the auto-control will act like a type of Maxwell demon, which is just a type of internal interactions. The demon may be a permeable membrane. Ordering is the formation of structure through the self-organization from a disordered state.

Ashby pointed out that two substances such as ammonia and hydrogen in a gaseous state can be mixed to form a solid [39]. Similarly, about twenty different types of amino acids present in microorganisms can gather together to form a new reproductive process. It is commonly understood that solids are more orderly than gases, such that the entropy of a solid should be less than the entropy of the same material in its gaseous state. Microorganisms should likewise represent a more orderly state than the amino acids from which they are formed. Yet in chemistry, the Belousov-Zhabotinski reaction shows a period change that occurs automatically, at least during specific time intervals.

In a biological self-organizing process, some isolated systems may spontaneously proceed toward an orderly state. Prigogine and Stengers have discussed such a case [40]: Under particular circumstance, such as when Dictyostelium discoideum experiences a lack of nutrition, solitary cells will spontaneously unite to form a larger cell cluster. In such a case, the cells and nutrition-liquid together may be regarded as an isolated system. Jantsch pointed out [41] that when different types of sponge and water are mixed within a uniform suspension, they rest for a few hours and then automatically separate into different types. More interestingly, when a small hydra is cut into its individual cells, the individual cells spontaneously evolve to form cell-clusters. Some cell clusters are malformations, but other cell clusters will eventually become a normal hydra.

We proposed nonlinear whole biology [42-44], which demonstrates that life itself is in an aspect of chaos in nature [44]. Based on the principles of nonlinear whole biology and the loop quantum theory, we researched a new method of quantum gravity for protein folding, and obtained four new approximate solutions and researched three possible origins for protein folding [43,44]. We have also studied a system of nonlinear whole medicine.

In a word, biology is a wide region for research of decrease of entropy in various isolated systems.

**3. Thought field and possible decrease of entropy in physiology and psychology**

Further, we research possible decrease of entropy in physiology, psychology and so on.

In physiology, increased metabolism and an emotional state of being upset should be characterized by larger entropy. Conversely, decreased metabolism and easy conscience or a calm and good-natured emotional state should be characterized by smaller entropy. The immunity of an organism increases for positive emotions, but metabolism and body-temperature show a remarkable increase for a nervous state and negative emotions increase an organism susceptibility to various diseases.

Anomalous cognition (AC) is defined as a form of information transfer in which all known sensorial stimuli are absent. Lantz, et al., reported to test sender condition and target type in AC experiment [45]. There is a difference between static and dynamic target material. Entropy is defined as a measure of uncertainty or lack of information about a system. The data from both of these studies were analyzed with regard to the gradient of Shannon entropy of the targets. May, et al., were able to compute the entropy and its mathematical gradient for each target in these experiments [46]. The AC was better when targets underwent massive changes in energy or entropy in a very short period of time. In addition, dynamic targets produced better results in the ganzfeld than did static targets a result that is suggestive of changes of entropy.

Based on experiments and the theoretical principles of modern science, we previously proposed a simple model of the thought field [47,48]. The thought field is emitted when a living being has an idea or thought. The energy of this field is directly proportional to a frequency:

$$E = H\nu. \qquad (21)$$

The frequency of the thought field can be strong or weak as well as concentric or disorderly. Different combinations of these four factors determine the various functional states displayed by any given person [47,48]. A strong concentric thought field corresponds to a functional state that is usually associated with the paranormal. This state includes Qigong and a special function, and is possibly analogous to the atom laser. The propagation of Qigong inside body is analogous to superfluid, which corresponds to lower entropy. The equation of superfluid has the soliton solution, which may describe the propagation. A possibility applied to earthquake prediction is searched [49].

In modern physics, scientists recognize only four fundamental interactions: the gravitational and electromagnetic fields whose action distances are infinite and thus represented as long-range forces, and the strong and weak fields whose action distances are extremely small and thus represented as short-range forces. New research has shown that the thought field has a mid-range of interaction. Its action distance is neither infinite nor very short and its strength is also mid-range when compared to the four known fundamental interactions. The thought field, which is related quantum teleportation [50], and the entangled state are possibly the new interactions [49,51,52].

Modern neuroscience believes that when the brain is stimulated by certain outside signals, it will produce induced electric potential, whose scale is related to the degree of attention. When the signals are stronger the potential is higher. Conversely, the potential is lower when the signals are weaker. This behavior is analogous to the thought field.

We discussed the quantitative aspects of the thought field as related to the extended quantum theory [50,53], in which the coefficient H in Eq.(21) may differ for various general cases. From this we may obtain the anthropic principle exactly [53,50]. We discussed the relation on the thought field and nonlinear whole biology [43].

**4.Possible decrease of entropy in some practices and social systems**

Lumsden-Cook considered emotion as a possible factor in mind-matter/PK interactions [54]. Two studies were conducted that examined how emotional states of anger and elation mediated the outputs of Random Even Generator (REG). These studies provide some support for the idea that affective states might influence REG activity. He also presented the results of two experiments investigating how emotional states can influence micro-psychokinetic functioning that may facilitate mind-matter interactions [55].

Landsberg defined disorder as the entropy normalized to the maximum entropy, which is that of the equiprobable distribution, corresponding to a completely random system. A living being as a whole represents an extremely orderly state of being and must be an open system for long-time. A living death represents a transformation to a state of total disorder, while sickness is a state of local disorder and a state of recovering from sickness is marked by a return to the higher order of health. The order parameters are thus health targets. But, for a short-time Qigong and some states attained during religious practices, for example, Buddhist and Taoist meditation, may be considered isolated systems that are characterized by decreases of entropy.

In the thought field different functional states are determined by the frequency. Such the thought field may interact with other oscillations. Strong or weak frequency, and concentric or disorderly frequency can be expressed by entropy and strength. This is namely thermodynamics in psychology.

In fact, Qigong inside body, which is a thought field with self-interactions, and some natural therapies have could often decrease entropy by internal self-control for a person as an isolated system. Qigong require that one must be calm and good-natured, which is a more order state. The clinical practices show that Qigong may reduce human metabolism. It expresses that human idea or thought oneself reduce entropy, and further achieve an ordering goal which may cure the sickness and build up human strength. The electroencephalogram (EEG) of Qigong emerges slow wave of children, which shows that electromagnetic activity of brain cells is high order.

In Buddhism everyone should face everyday in a happy mood with thanks to Nature. A harmonious unification between two activities of human body and spirit can only consist completely of a normal sound activity.

The social thermodynamics is a part of the social physics (sociophysics) [56,57]. Ilya Prigogine proposed order through fluctuation for self-organization and social system. Lepkowski discussed the social thermodynamics of Ilya Prigogine [58]. Reed and Harvey discussed complexity and realism in the social sciences, and critical philosophy and non-equilibrium thermodynamics [59]. E.L.Khalil (1995) studied nonlinear thermodynamics and social science modeling. Scafetta, et al., studied concretely the thermodynamics of social processes for the teen birth phenomenon [60]. Zagreb researched approach to a quantitative description of social systems based on thermodynamic formalism [61].

Further, Rifkin and Howard proposed entropy as a new world view [62]. J.L.R.Proops (1987) discussed entropy, information and confusion in the social sciences. K.D. Bailey stated social entropy theory (1990), and its application of nonequilibrium thermodynamics in human ecology (1993). Then he discussed living systems theory and social entropy theory [63]. G.A.Swanson, K.D.Bailey and J.G.Miller (1997) discussed entropy, social entropy and money. Balch (1997) discussed social entropy as a new metric for learning multi-robot teams, and researched the hierarchic social entropy as an information theoretic measure of robot group diversity [64]. S.Robinson, A.Cattaneo and M.El-Said (2001) studied updating and estimating a social accounting matrix using cross entropy methods. K.Mayumi and M.Giampietro discussed entropy in ecological economics in book <Modelling in Ecological Economics> (2004). Stepanic, et al., described social systems using social free energy and social entropy [65].

Finally, we research possible decrease of entropy in some social systems. The certain commodity and the two parties of business construct an isolated system in economics, but it will not usually tend to a disorder state, a price will reach stable through some arrange (i.e., internal interaction), and the system will become order. It is well known that if some people exchange views (i.e., internal interaction in isolated system), new thought and information will be produced. In some social systems, even if the constitutes and circumstance are the same, as long as the structure is optimized, then the system will become more order. In a word, the evolution of Earth and biology, and human history are always not a disorder process. Of course, human aftertime is also not a nice heaven without misery. Human existence and development, order or disorder, determine on the human internal relations, and on the interactions between humanity and other entirnment.